\documentclass[10pt,pra,twocolumn,superscriptaddress,showpacs,floatfix]{revtex4}

\usepackage{graphicx}
\usepackage{amssymb}
\usepackage{times}
\usepackage{amsmath}
\usepackage{amsthm}
\usepackage{epstopdf}
\usepackage{verbatim}
\usepackage{hyperref}

\input epsf.tex
\usepackage{graphicx}
\usepackage{amsthm}

\begin{document}

\title{Security bound of continuous-variable measurement-device-independent quantum key distribution with imperfect phase reference calibration}

\author{Hong-Xin Ma}\affiliation
 {Henan Key Laboratory of Quantum Information and Cryptography, Zhengzhou Information Science and Technology Institute, Zhengzhou, Henan 450001, China}
\affiliation{Synergetic Innovation Center of Quantum Information and Quantum Physics, University of Science and Technology of China, Hefei, Anhui 230026, China}
\affiliation
 {State Key Laboratory of Advanced Optical Communication Systems and Networks and Center of Quantum Information Sensing and Processing, Shanghai Jiao Tong University, Shanghai 200240, China
}

\author{Peng Huang}\thanks{Corresponding author: huang.peng@sjtu.edu.cn}
\affiliation
 {State Key Laboratory of Advanced Optical Communication Systems and Networks and Center of Quantum Information Sensing and Processing, Shanghai Jiao Tong University, Shanghai 200240, China
}

\author{Tao Wang}\affiliation
 {State Key Laboratory of Advanced Optical Communication Systems and Networks and Center of Quantum Information Sensing and Processing, Shanghai Jiao Tong University, Shanghai 200240, China
}

\author{Dong-Yun Bai}\affiliation
 {State Key Laboratory of Advanced Optical Communication Systems and Networks and Center of Quantum Information Sensing and Processing, Shanghai Jiao Tong University, Shanghai 200240, China
}

\author{Shi-Yu Wang}\affiliation
 {State Key Laboratory of Advanced Optical Communication Systems and Networks and Center of Quantum Information Sensing and Processing, Shanghai Jiao Tong University, Shanghai 200240, China
}

\author{Wan-Su Bao}\affiliation{Henan Key Laboratory of Quantum Information and Cryptography, Zhengzhou Information Science and Technology Institute, Zhengzhou, Henan 450001, China}
\affiliation{Synergetic Innovation Center of Quantum Information and Quantum Physics, University of Science and Technology of China, Hefei, Anhui 230026, China}

\author{Gui-Hua Zeng}\affiliation
 {State Key Laboratory of Advanced Optical Communication Systems and Networks and Center of Quantum Information Sensing and Processing, Shanghai Jiao Tong University, Shanghai 200240, China
}

\date{\today}

\begin{abstract}
Phase reference calibration is a necessary procedure in practical continuous-variable measurement-device-independent quantum key distribution (CV-MDI-QKD) for the need of Bell-State Measurement (BSM). However, the phase reference calibration may become imperfect in practical applications.
We explored the practical security of CV-MDI-QKD with imperfect phase reference calibration under realistic conditions of lossy and noisy quantum channel.
Specifically, a comprehensive framework is developed to
model and characterize the imperfection of practical phase reference calibration operation, which is mainly caused by the non-synchronization of two remote lasers in senders.
Security analysis shows that the imperfect phase reference calibration has significant side effects on the performance and security of the CV-MDI-QKD protocol.
A tight security bound to excess noise introduced by imperfect phase reference calibration is derived for reverse reconciliation against arbitrary collective attacks in the asymptotic limit, and the tolerance of the CV-MDI-QKD protocol to this excess noise is also obtained.
This security analysis framework can eliminate the security hazards caused by imperfect phase reference calibration without changing the existing CV-MDI-QKD system structure.




\end{abstract}


\maketitle

\section{Introduction}\label{Intr}
Quantum key distribution (QKD) \cite{NG02} allows two distant authenticated users, Alice and Bob, to establish secure key through untrusted envrionment, which is based on the principles of quantum mechanics.
There are mainly two categories of QKD: discrete-variable (DV)QKD protocols \cite{BB84,E91} and continuous-variable (CV) QKD protocols \cite{TC99,GG02,GG03,CS12}.
CVQKD utilizes the quadrature components of quantum states to distribute the secure key, which has unique potentials of being compatible with standard telecommunication systems and no request on single-photon detectors.
Furthermore, CVQKD allow users to approximate the PLOB bound \cite{PLOB17}, which depicts the ultimate limit of repeater-less communication.

Theoretically, the Gaussian-modulated CVQKD protocol using coherent states \cite{GG02} has been proved to be secure against arbitrary collective attacks \cite{COL06} and coherent attacks \cite{COR09}, even with finite-size regime \cite{FIN13,FIN17} and composable security \cite{COM15} taken into account.
Experimentally, this protocol has been proved to be feasible both in laboratory \cite{GG03,LAB13} and field tests \cite{FIE16}.
The Gaussian-modulated CVQKD protocol has extended the secure transmission over 100 km optical fiber in the laboratory \cite{150KM}, which shows its potential of applying in metropolitan quantum networks.

The security analysis of CVQKD relies on some ideal assumptions, which are hard to satisfy in practice \cite{IMP00,IMP10}.
These deviations will bring specific security vulnerabilities to CVQKD system, and the eavesdroppers can utilize this imperfection to implement attack strategies, such as local oscillator fluctuation attack \cite{LOF13}, calibration attack \cite{CAL13}, wavelength attack \cite{WAV13}, detector  saturation attack \cite{SAT16}. Obviously, most of these attack strategies mainly focus on the imperfect detectors.
In order to remove these attacks, one solution is to find and describe these security vulnerabilities, and then propose corresponding countermeasures.
But characterizing all vulnerabilities is quite difficult, and the countermeasures will increase the complexity of the system.

Inspired by the idea of entanglement swapping, measurement-device-independent (MDI) QKD has been proposed by two groups \cite{SL12,LHK12} independently, which can eliminate all side-channel attacks on detectors.
Continuous-variable MDI-QKD (CV-MDI-QKD) has been proposed and verified both theoretically and experimentally \cite{STN15}. Some theoretical schemes of CV-MDI-QKD have been put forward one after another in the same period\cite{MXC14,LZY14,ZYC14,COT15}.
In the theoretic research of CV-MDI-QKD, some tremendous results have been achieved in recent years \cite{XZ17,PP17,CLU18,CLU218,CZY18,ZYJ18,MHX18,BAI19}.
In CV-MDI-QKD protocols, Alice and Bob are both senders, and measurement operations are performed by an untrustworthy third party, Charlie. Charlie performs Bell-State Measurement (BSM) based on signals sent by Alice and Bob, where the measurement result is communicated publicly and used for generating the secure keys.
Since measurement operations are performed by untrusted terminal, the security of CV-MDI-QKD does not depend on the detectors. In other words, CV-MDI-QKD can eliminate all side-channel attacks against detectors, whether known or unknown.

In practical system of CV-MDI-QKD, the light sources of Alice and Bob are mutually independent. Therefore, the initial optical pulses they emit are also independent of each other and may not stay in the same phase reference frame.
For the need of BSM, we need to calibrate the phase reference frames between Alice, Bob and Charlie \cite{STN15}.
The basic idea of phase reference calibration in CV-MDI-QKD is described as follows: first, we measure the phase difference between the local oscillator pulses emitted by Alice and Bob.
Then, we take relative phase estimation and correction, adding the phase difference to one side's quantum signal pulse. After these operations, Alice and Bob's quantum signal pulses are stay in the same phase reference frame, and Charlie carries out BSM based on this unified phase reference frame.

Obviously, phase reference calibration is of vital importance for the construction of experimental framework for CV-MDI-QKD.
Unfortunately, in practical implementation, the phase reference calibration operation is not as perfect as theory.
Due to the non-synchronization of two independent lasers in Alice and Bob's sides, which are mainly caused by the separate spectral linewidths of two lasers , and the uncertainty of the channel and detection environment, the practical phase reference calibration operation will become imperfect.
If the imperfection is not taken into account in security analysis, the security key rate obtained will be higher than the actual value, which may lead to security hazards.
For the accuracy of security analysis, in other words, in order to get a tighter bound of security key rate, we need to precisely characterize the impact of imperfect phase reference calibration in security analysis process.

Some latest breakthroughs \cite{IPC18,IPC19} overcome the non-ideality brought about by the practical phase reference calibration to a certain extent through the new optical path design, which simplify the phase reference calibration process.
However, these schemes may increase the complexity of other aspects of the system, such as detection, optical path and so on. In addition, these schemes may also introduce additional excess noise, such as the phase noise between signal pulse and reference pulse.
In this paper, we choose to deal with this problem from another point of view, that is, to quantitatively characterize the imperfection of practical phase reference calibration operation through reasonable modeling, which develops a comprehensive security framework of CV-MDI-QKD protocol with imperfect phase reference calibration. The exact formula for calculating excess noise caused by the imperfect phase reference calibration is obtained, and then a more compact and accurate security key rate is derived under arbitrary collective attacks. Based on this, we can qualitatively and quantitatively analyze the impact of imperfect phase reference calibration on the performance and security of CV-MDI-QKD protocol.
This security analysis framework can eliminate the security hazards caused by imperfect phase reference calibration without changing the existing CV-MDI-QKD system structure.

\begin{figure}[!h]\center
\resizebox{8.5cm}{!}{
\includegraphics{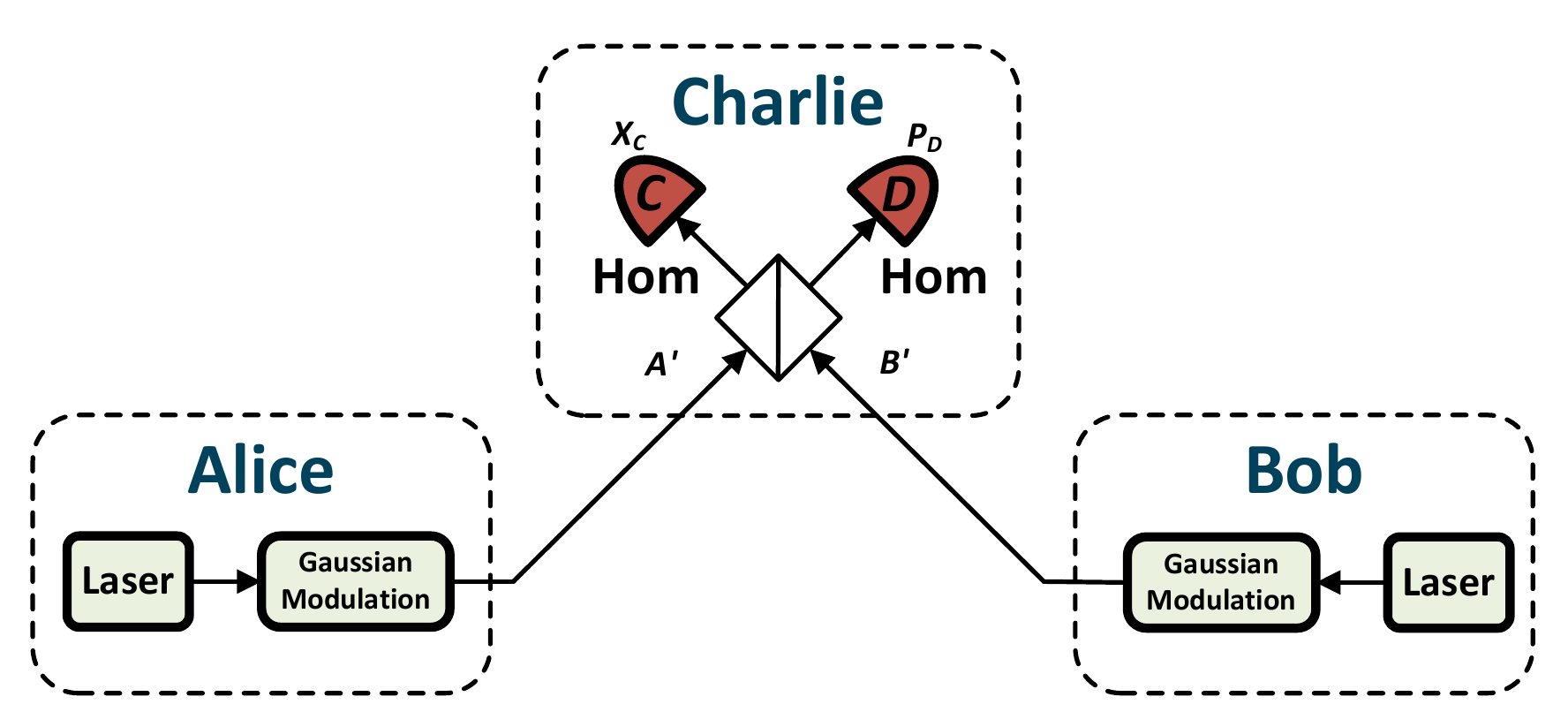}}
\label{fig1}
\caption{(Color online). PM version of the CV-MDI-QKD protocol. Hom is homodyne detection.}\label{fig:1}
\end{figure}

The remainder of this paper is structured as follows. In Sec.~\ref{IPC}, we first review the structure of CV-MDI-QKD protocol, then introduce phase reference calibration in CV-MDI-QKD protocol and develop a comprehensive framework to obtain the excess noise introduced by imperfect phase reference calibration. In Sec.~\ref{Cal}, we derive the secret key rate of the CV-MDI-QKD protocol with imperfect phase reference calibration, which is more precise and compact than the original one. In Sec.~\ref{PerD},we give the numerical simulation and performance analysis. Conclusion and discussions are drawn in Sec.~\ref{Con}.

\section{CV-MDI-QKD protocol with imperfect phase reference calibration}\label{IPC}  

In this section, we first review the CV-MDI-QKD protocol, especially the prepare-and-measure (PM) version. Then, we introduce the phase reference calibration operation in CV-MDI-QKD protocol and its imperfection in practical implementation. On the basis of these reviews, we describe and calculate the excess noise caused by imperfect phase reference calibration by precise modeling.

\subsection{CV-MDI-QKD Protocol}

The construction of CV-MDI-QKD protocol is illustrated in Fig.~\ref{fig:1}, which is based on the PM version. The main steps of PM version can be depicted as follows:

Step 1: Alice and Bob each prepare coherent states and send them to third-party Charlie through two different quantum channels with length \(L_{AC}\) and \(L_{BC}\), respectively. The coherent state prepared by Alice is \(\left| {{x_A} + i{p_A}} \right\rangle \), where \(x_A\) and \(p_A\) are Gaussian distributed with modulation variance \(V_{AM}\). The coherent state prepared by Bob is \(\left| {{x_B} + i{p_B}} \right\rangle \), where \(x_B\) and \(p_B\) are Gaussian distributed with modulation variance \(V_{BM}\).

Step 2: Charlie performs BSM by interfering the two incoming coherent states on a beam splitter and obtaining two output modes \(C\) and \(D\). Then, Charlie use two homodyne detections to measure the \(x\) quadrature of mode \(C\) and \(p\) quadrature of mode \(D\) and announced the measurement results \(\left\{X_C,P_D\right\}\) publicly.

Step 3: After receiveing Charlie's measurement results, Alice keeps her data unchanged, where \(X_A=x_A, P_A=p_A\), while Bob modifies his data to \(X_B=x_B+\kappa X_C\), \(P_B=p_B-\kappa P_D\). \(\kappa \) is an optimization parameter associated with quantum channel loss.

\begin{figure}[!h]\center
\resizebox{8.5cm}{!}{
\includegraphics{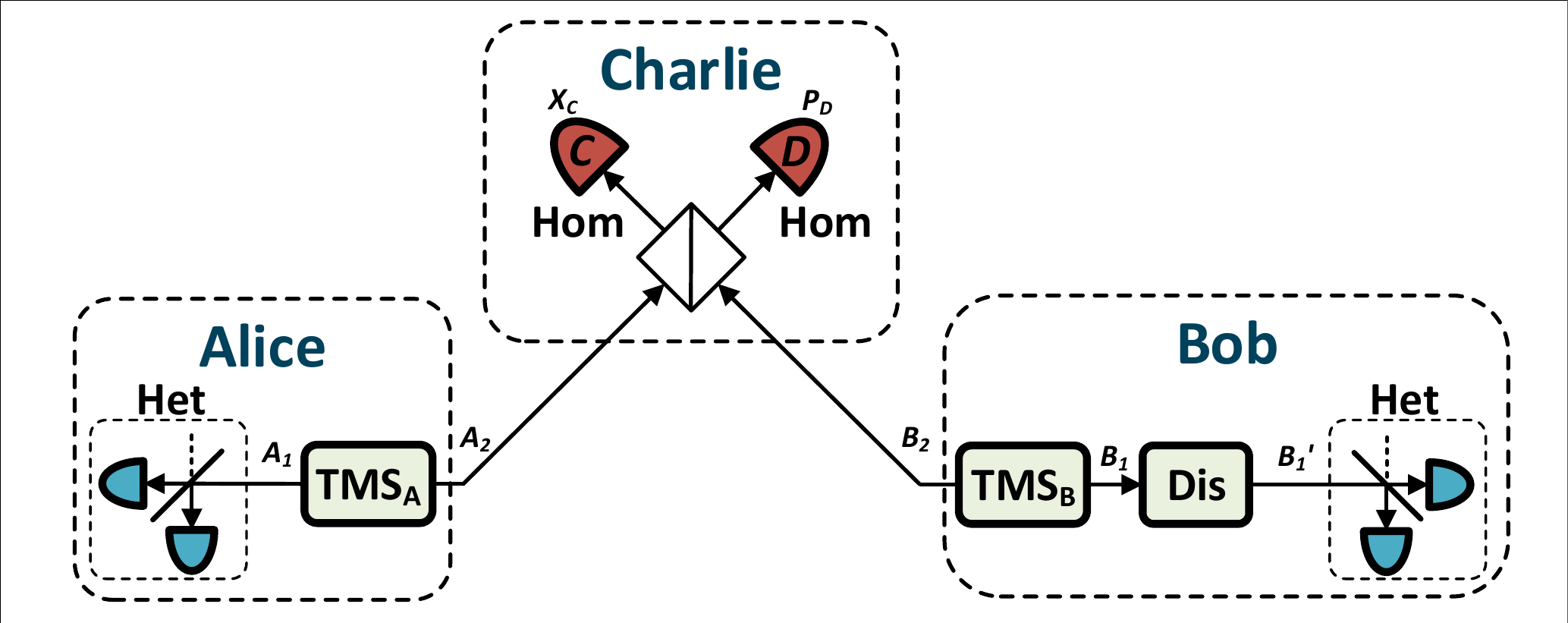}}
\label{fig1}
\caption{(Color online). EB version of the CV-MDI-QKD protocol. Het is heterodyne detection, Dis is displacement operation, \(\rm{TMS_A}\) and \(\rm{TMS_B}\) are two-mode squeezed states.}\label{fig:2}
\end{figure}

Step 4:  Alice and Bob extract a string of secret key after carrying out parameter estimation, information reconciliation and privacy amplification steps through an authenticated public channel.

In the equivalent entanglement-based (EB) version, which is shown in Fig.~\ref{fig:2}, Alice and Bob prepare two-mode squeezed states independently and each send one mode to to Charlie for BSM. After Charlie announced the measurement results, Bob displaces his retained mode according to the measurement results, where the gain of the displacement operation is \(g\), while Alice keeps her mode unchanged. Then, Alice and Bob measure their modes to obtain the raw data. After the date post-processing, Alice and Bob obtain the final secret keys.

Before this series of steps, Alice and Bob implement the phase reference calibration by  measuring the phase difference between the local oscillator pulses emitted by Alice and Bob, which makes sure that the prepared coherent states (or two-mode squeezed states) of Alice and Bob stay in the same phase reference frame.

\subsection{Phase reference calibration in CV-MDI-QKD}

This subsection mainly discusses the definition and operation of phase reference calibration between Alice, Bob and Charlie in CV-MDI-QKD protocol.

Practically, local oscillator pulses, as the phase reference light of signal pulse, can be a strong classical light.
Therefore, by interfering two classical local oscillator lights on a beam splitter, the phase difference of the two local oscillator pulses can be measured by measuring the intensity of one output beam with photon detector.

\begin{figure}[!h]\center
\resizebox{9cm}{!}{
\includegraphics{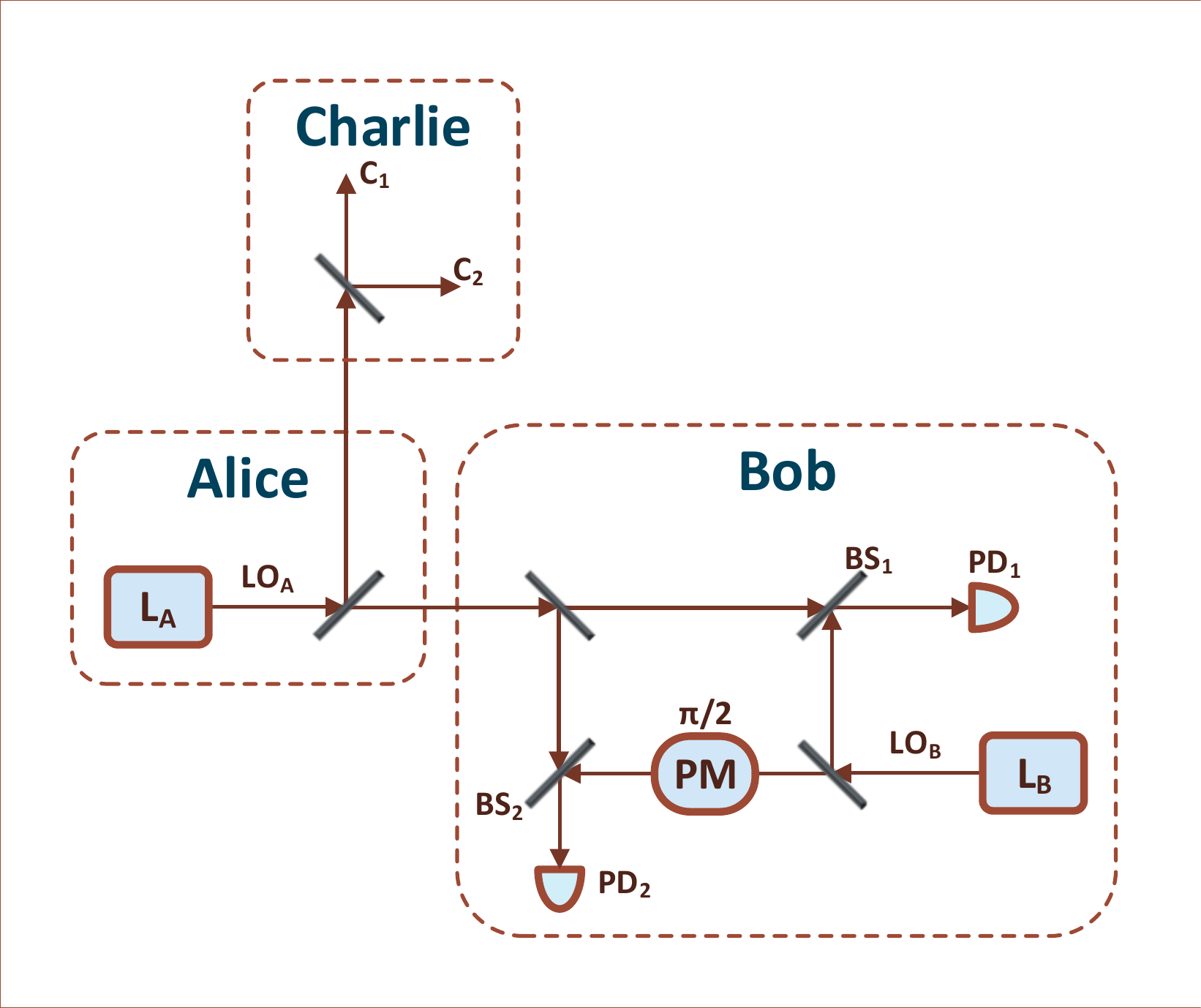}}
\caption{(Color online). Schematic structure of measuring the phase difference between the local oscillators sending by Alice and Bob in CV-MDI-QKD protocol.
PM is phase modulator. \(\rm{L_A}\) is the laser in Alice's side, \(\rm{L_B}\) is the laser in Bob's side. \(\rm{LO_A}\) and \(\rm{LO_B}\) are local oscillator pulses. \(\rm{PD_1}\) and \(\rm{PD_2}\) are photo detectors. \(\rm{C_1}\) and\(\rm{C_2}\) are the reference lights of two balanced homodyne detectors for BSM. \(\rm{BS_1}\) and \(\rm{BS_2}\) are beam-splitters. The ratio of all the beam-splitters is 50:50.
}\label{fig:3}
\end{figure}

We assume that the measurement of phase difference and phase reference calibration are performed by Bob. Alice sends her local oscillator pulse to the untrusted third part Charlie .
The schematic diagram of apparatus for measuring the phase difference of the local oscillator pulses is given in Fig.~\ref{fig:3}.
Alice divides its local oscillator pulse \(\rm{LO_A}\) into two beams, one sent to Charlie and the other one sent to Bob.
Charlie divides the received beam into two beams as the reference lights of two balanced homodyne detectors for BSM.
After receiving the local oscillator pulse sent by Alice, Bob divides the received local oscillator pulse and his own local oscillator pulse \(\rm{LO_B}\) into two beams respectively, and interferences these beams through \(\rm{BS_1}\) and \(\rm{BS_2}\). In order to measure the phase difference accurately, \(\pi/2\) phase has been added to one of the local oscillator beams.
Then, the phase difference of the two local oscillator pulses can be obtained by measuring the output interference intensity of one port of \(\rm{BS_1}\) and \(\rm{BS_2}\) respectively with \(\rm{PD_1}\) and \(\rm{PD_2}\).

The local oscillator pulses \(\rm{LO_A}\) and \(\rm{LO_B}\) can be denoted as \(\alpha _{LO}^A{e^{i{\theta _A}}}\) and \(\alpha _{LO}^B{e^{i{\theta  _B}}}\) respectively. \(|\alpha _{LO}^A|\) and \(|\alpha _{LO}^B|\) are the amplitude of each local oscillator pulses. \(\theta_A\) and \(\theta _B\) are the phase of \(\rm{LO_A}\) and \(\rm{LO_B}\), respectively.
We suppose \(\alpha _{LO}^A=\alpha _{LO}^B=\alpha_{LO}\).
After local oscillator pulses interferes on these beam splitters, the amplitude of the light measured by PD1 can be expressed as
\begin{equation}
\begin{array}{lll}
{\beta _1} &= \frac{1}{{\sqrt 2 }}\left( {{\alpha _{LO}}{e^{i{\theta _A}}} + {\alpha _{LO}}{e^{i{\theta _B}}}} \right)\\ \\
 &= \sqrt 2 {\alpha _{LO}}{e^{\frac{{i\left( {{\theta _A} + {\theta _B}} \right)}}{2}}}\cos \left( {\frac{{{\theta _A} - {\theta _B}}}{2}} \right),
\end{array}
\end{equation}
then the intensity of the light measured by PD1 can be calculated as
\begin{equation}
\begin{array}{lll}
{\left| {{\beta _1}} \right|^2} &= 2{\left| {{\alpha _{LO}}} \right|^2}{\cos ^2}\left( {\frac{{{\theta _A} - {\theta _B}}}{2}} \right)\\ \\
 &= {\left| {{\alpha _{LO}}} \right|^2}\left[ {1 + \cos ({\theta _A} - {\theta _B})} \right].
\end{array}
\end{equation}
Similarly, the intensity of the light measured by PD2 is obtained as
\begin{equation}
\begin{array}{lll}
{\left| {{\beta _2}} \right|^2} &= {\left| {{\alpha _{LO}}} \right|^2}\left[ {1 + \cos ({\theta _A} - {\theta _B} - \pi /2)} \right]\\ \\
 &= {\left| {{\alpha _{LO}}} \right|^2}\left[ {1 + \sin ({\theta _A} - {\theta _B})} \right].
\end{array}
\end{equation}
According Eq.(2) and Eq.(3), we can obtain the phase difference between Alice's and Bob's local oscillator, \(\varphi_{cal}\), which is calculated as
\begin{equation}
{\varphi _{cal}} = {\theta _A} - {\theta  _B}.
\end{equation}

After the phase reference calibration operation, the correlation between \(\left( {{X^A_{LO}},{P^A_{LO}}} \right)\) and \(\left( {{X^B_{LO}},{P^B_{LO}}} \right)\) can be obtained by

\begin{equation}
\begin{array}{l}
{X^B_{LO}} = {X^A_{LO}}\cos {\varphi _{cal}} - {P^A_{LO}}\sin {\varphi _{cal}},\\ \\
{P^B_{LO}} =  {X^A_{LO}}\sin {\varphi _{cal}} + {P^A_{LO}}\cos {\varphi _{cal}}
\end{array}
\end{equation}

Assuming the Alice's local oscillator has a zero-phase angle, which means \(P^A_{LO}=0\), the expression of \(\varphi_{cal}\) can be obtained as
\begin{equation}
{\varphi _{cal}} = {\tan ^{ - 1}}\left( {{P^B_{LO}}/{X^B_{LO}}} \right).
\end{equation}

Relatively to local oscillator pulses, the initial quantum signal pulses modulated by Alice and Bob can be expressed as \(\alpha _S^A{e^{i\left( {{\theta _A} + {\theta _{AM}}} \right)}}\) and \(\alpha _S^B{e^{i\left( {{\theta _B} + {\theta _{BM}}} \right)}}\) respectively. \(\alpha _S^A\) and \(\alpha _S^B\) are the intensities of their respective signal pulses,  \(\theta_{AM}\)  and \(\theta_{BM}\) are their initial modulated phases, respectively.
Based on the phase difference \(\varphi_{cal}\) between Alice's and Bob's local oscillator pulses, when Bob modulates his quantum signal pulses, the phase difference \(\varphi_{cal}\) and the initial modulated phase \(\theta_{BM}\) should be added as the modulated phase of his ultimate modulated quantum signal pulse, which can be expressed as
\begin{equation}
\alpha _S^B{e^{i\left( {{\theta _B} + {\theta _{BM}} + {\varphi _{cal}}} \right)}} = \alpha _S^B{e^{i\left( {{\theta _A} + {\theta _{BM}}} \right)}}.
\end{equation}
Obviously, Bob's ultimate modulated quantum signal pulse is defined in Alice's quantum signal modulation reference frame. At this time, Alice and Bob's quantum signal pulses share the same phase reference frame.

\subsection{Excess noise introduced by imperfect phase reference calibration}

Theoretically, after local oscillator reference quadrature measurement, relative phase estimation and correction,
Alice and Bob's quantum signal pulses are expected to stay in the same phase reference frame with the phase difference \({\varphi _{cal}} \).
However, in practice, the phase reference calibration operation is not as as perfect as in theory, and the estimator \({{\hat \varphi }_{cal}}\) always has estimation error, which leads to excess noise.
In the case of Gaussian-modulated protocol, we assume the excess noise introduced by imperfect phase reference calibration is Gaussian, which is similar with the specific phase noise denoted in Ref. \cite{SFCV,PN}, can be written as
\begin{equation}
{\varepsilon _{prc}} = 2V_M(1 - {e^{ - {V_{prc}}/2}}),
\end{equation}
where \(V_M=V_{AM}=V_{BM}\) is both the modulation variance of Alice and Bob, \(V_{prc}\) is the variance of the excess noise introduced by imperfect phase reference calibration, which is expressed as \cite{SELF,SELF2}
\begin{equation}
{V_{prc}} = {\mathop{\rm var}} ({\varphi _{cal}} - {{\hat \varphi }_{cal}}).
\end{equation}
Assuming that the laser in Alice's side, \(\rm{L_A}\), have spectral linewidth \(\Delta {\nu _A}\), and the laser in Bob's side, \(\rm{L_B}\), have spectral linewidth \(\Delta {\nu _B}\). Both lasers are centered around the same optical frequency.
\(f\) is the repetition rate of the system.
The excess noise \(V_{prc}\) is constituted by three terms
\begin{equation}
{V_{prc}} = V_{laser}+V_{measure}+V_{path}.
\end{equation}
The term \(V_{laser}\) represents the variance of the relative phase drift between two free-running lasers \(\rm{L_A}\) and \(\rm{L_B}\), which can be obtained as
\begin{equation}
{V_{laser}} = \frac{2\pi}{f}{(\Delta {\nu _A} + \Delta {\nu _B})}.
\end{equation}
Obviously, \({V_{laser}}\) is caused by the fact that the pulses of \(L_A\) and \(L_B\) are non-synchronization, which mainly leads by the separate spectral linewidths of two lasers. In the specific system of CV-MDI-QKD protocol, \(V_{laser}\) is a fixed parameter.

The term \(V_{measure}\) corresponds to the noise that caused by the measurement error of the local oscillator phase. In CV-MDI-QKD protocol, \(V_{measure}\) can be expressed as
\begin{equation}
\begin{array}{lll}
V_{measure}&=\frac{\chi_A+1}{{|\alpha^A_{LO}|}^2}+\frac{\chi_B+1}{{|\alpha^B_{LO}|}^2}
\\ \\
&=\frac{\chi_A+\chi_B+2}{{|\alpha_{LO}|}^2},
\end{array}
\end{equation}
where \(\chi_A\) is the total noise imposed on the local oscillator \(\rm{LO_A}\), which is send by Alice to Charlie, and \(\chi_B\) is the total noise imposed on the local oscillator \(\rm{LO_B}\), which is send by Bob to Charlie. \(|\alpha^A_{LO}|\) and \(|\alpha^B_{LO}|\) are the amplitude of the local oscillators \(\rm{LO_A}\) and \(\rm{LO_B}\) respectively, and \(\alpha _{LO}^A=\alpha _{LO}^B=\alpha_{LO}\). \(\chi_A\) and \(\chi_B\) are defined in Eq. (15).

The term \(V_{path}\) represent the relative phase drift which caused by the accumulation of the phase difference between the quantum signal pulse and the local oscillator pulse. Practically, it is caused by the different optical path lengths between two kind pulses.
In CV-MDI-QKD protocol, the quantum signal pulse and the local oscillator pulse transmit through the same optical path each for Alice and Bob. Thus we have \(V_{path}=0\), and the excess noise \(V_{prc}\) is caused by two major components: \({V_{prc}} = V_{laser}+V_{measure}\).

When the deviation of \({{\hat \varphi }_{cal}}\) is quite small, \(V_{prc}\) keeps in a relatively low range. Under this condition, the excess noise introduced by imperfect phase reference calibration can be approximated as \cite{SELF}

\begin{equation}
\begin{array}{lll}
{\varepsilon _{prc}}&=&V_M{V_{prc}}
\\ \\
&=& 2\pi \frac{V_M(\Delta {\nu _A} + \Delta {\nu _B})}{f}+\frac{V_M(\chi_A+\chi_B+2)}{{|\alpha_{LO}|}^2}.
\end{array}
\end{equation}



We denote the transmittance of the quantum channel between Alice (Bob) and Charlie is \(T_A\) (\(T_B\)), and both quantum channel losses are \(l\) = 0.2 dB/km, then the transmittance can be given as
\({T_A} = {10^{\frac{{ - l {L_{AC}}}}{{10}}}}\), \({T_B} = {10^{\frac{{ - l {L_{BC}}}}{{10}}}}\).
The excess noise introduced by two separate quantum channels are \(\varepsilon_A\) and \(\varepsilon_B\), respectively.
\(\varepsilon_c\) is the equivalent excess noise introduced by all quantum channels, which is obtained as
\begin{equation}
\begin{array}{lll}
\varepsilon_{c}  =&1+{\chi _A}+\frac{{{T_B}}}{{{T_A}}}{\left( {{\chi _B} - 1} \right)}
\\ \\
& +\frac{{{T_B}}}{{{T_A}}}{{\left( {\sqrt {\frac{2}{{{T_B}{g^2}}}}\sqrt {{V_B}-1}  - \sqrt {{V_B} + 1} } \right)}^2},
\end{array}
\end{equation}
where \(V_B=V_{BM}+1\), \(g\) is the amplification coefficient of the Bob's displacement in EB version, and
\begin{equation}
{\chi _A} = \frac{1}{{{\eta_A}}} - 1 + {\varepsilon _A}, {\chi _B} = \frac{1}{{{\eta_B}}} - 1 + {\varepsilon _B}.
\end{equation}
As \(g\) is an optimization parameter, we denote \({g^2} = \frac{{2({V_B-1}) }}{{{\eta_B}\left( {{V_B} + 1} \right)}}\) to minimize \(\varepsilon\). Then the optimized equivalent excess noise introduced by all quantum channels can be calculated as
\begin{equation}
\varepsilon_c  = \frac{{{T_B}}}{{{T_A}}}\left( {{\varepsilon _B} - 2} \right) + {\varepsilon _A} + \frac{2}{{{T_A}}}.
\end{equation}

We suppose the homodyne detectors in Charlie are ideal apparatuses, then the total added noise expressed in shot noise units is
\begin{equation}
\chi _{t}=\frac{1 }{\eta} -1+ \varepsilon_c+\varepsilon_{prc},
\end{equation}
where \(\eta = \frac{1}{2}{g^2}{T_A}\) is a normalized parameter associated with the total quantum channel transmittance \cite{LZY14}.

\section{Calculation of the secret key rate}\label{Cal}

In this section,  we will derive the secret key rate of the CV-MDI-QKD protocol against arbitrary collective attacks with considering the imperfection of practical  phase reference calibration operation.

In this paper, we mainly focus on the one-mode attack, where Eve takes entangling cloner attacks on each quantum channel independently. We should point out that this attack strategy is not the optimal one. The two-mode attack \cite{STN15}, where Eve takes correlated two-mode coherent Gaussian attack on two quantum channels by employing their interactions, is demonstrated to be the optimal attack strategy against the CV-MDI-QKD protocol. However, when the two quantum channels come from different directions, their correlation should be very weak, and it is extremely difficult for Eve to employ the correlation in practice. In this context, we approximately reduce the quantum channel of CV-MDI-QKD to one-mode channel, and one-mode attack can work efficiently. In addition, when \(\rm{TMS_B}\) and the displacement operation are regarded as manipulated by Eve, the EB version of CV-MDI-QKD protocol can be simplify to an equivalent one-way CVQKD protocol. Then we can use the secret key rate of equivalent one-way protocol to obtain the lower bound of the secret key rate of our protocol

Considering the lossy and noisy quantum channel and imperfection of practical phase reference calibration, the covariance matrix of \(\rho _{{A_1}{B'_1}}\) in EB version can be expressed as
\begin{equation}
\begin{array}{lll}
\gamma _{{A_1}{{B'}_1}} &= \left( {\begin{array}{*{20}{c}}
{a{{\rm{I}}_2}}&{c{\sigma _z}}\\
{}&{}\\
{c{\sigma _z}}&{b{{\rm{I}}_2}}
\end{array}} \right) \\ \\
&= \left( {\begin{array}{*{20}{c}}
{V{{\rm{I}}_2}}&{\sqrt {\eta(V^2-1)}{\sigma _z}}\\
{}&{}\\
{\sqrt {\eta(V^2-1)}{\sigma _z}}&{{\eta}\left( {V + {\chi _t}} \right){{\rm{I}}_2}}
\end{array}} \right),
\end{array}
\end{equation}
where \({\rm I}_2\) is \(2 \times 2\) identity matrix, \({\sigma _z}=diag(1,-1)\), \(V=V_A=V_B=V_M+1\).

The secret key rate of the CV-MDI-QKD protocol with imperfect phase reference calibration  under reverse reconciliation can be calculated as

\begin{equation}
{K_{prc}} = \beta {I_{AB}} - {\chi _{BE}},
\end{equation}
where \(\beta\) is the reconciliation efficiency, \(\chi _{BE}\) is the Holevo bound \cite{HOLEVO} which defines the maximum information available to Eve on Bob's key, \(I_{AB}\) is the mutual information between Alice and Bob,
which can be calculated by \cite{HOLEVO}
\begin{equation}
I_{AB} =2 \times \frac{1}{2} {\log _2}\left[ {\frac{{a + 1}}{{a + 1 - {c^2}/(b+1)}}} \right].
\end{equation}
The Holevo bound\(\chi _{BE}\) is given as
\begin{equation}
{\chi _{BE}} = S\left( {{\rho _{E}}} \right) - \int {d{m_B}p\left( {{m_B}} \right)} S\left( {\rho _{E}^{{m_B}}} \right),
\end{equation}
where \(S\) is the Von Neumann entropy of the quantum state \(\rho\), \(m_B\) represents the measurement of Bob,  \(p(m_B)\) is the probability density of the measurement, \(\rho _{E}^{{m_B}}\)  is Eve's state conditional on Bob's measurement result.
Based on the fact that \({\rho _{A_1}^{{m_B}}}\) is independent of \(m_B\) for Gaussian protocols, and Eve purifies the system \({A_1}{B'_1}\), \(\chi _{BE}\) can be obtained as
\begin{equation}
{\chi _{BE}} =S\left( {{\rho_{{A_1}{B'_1}}}} \right) - S\left( {\rho_{A_1}^{m_{B}}} \right),
\end{equation}
where \(S\left( {{\rho_{{A_1}{B'_1}}}} \right)\) is a function of the symplectic eigenvalues \(\lambda _{1,2}\) of \(\gamma_{{A_1}{{B'}_1}}\) characterizing the state \(\rho _{{A_1}{B'_1}}\), with the form
\begin{equation}
S\left( {{\rho_{{A_1}{B'_1}}}} \right) =G[(\lambda _1-1)/2]+G[(\lambda _2-1)/2],
\end{equation}
and \(S\left( {\rho_{A_1}^{m_{B}}} \right)\) is a function of the symplectic eigenvalues \(\lambda_3\) of \(\gamma_{{A_1}}^{{m_{B}}}\) characterizing the state \(\rho _{{A_1}}^{{m_{B}}}\), with the form
\begin{equation}
S\left( {\rho_{A_1}^{m_{B}}} \right)=G[(\lambda _3-1)/2],
\end{equation}
where the Von Neumann entropy
\begin{equation}
G\left( x \right) = \left( {x + 1} \right){\log _2}(x + 1) - x{\log _2}x.
\end{equation}
The symplectic eigenvalues \(\lambda_{1,2}\) can be calculated by
\begin{equation}
\lambda  _{1,2}^2 = \frac{1}{2}\left( {A \pm \sqrt {{A^2} - 4B^2} } \right),
\end{equation}
\begin{figure}[!h]\center
\resizebox{8.5cm}{!}{
\includegraphics{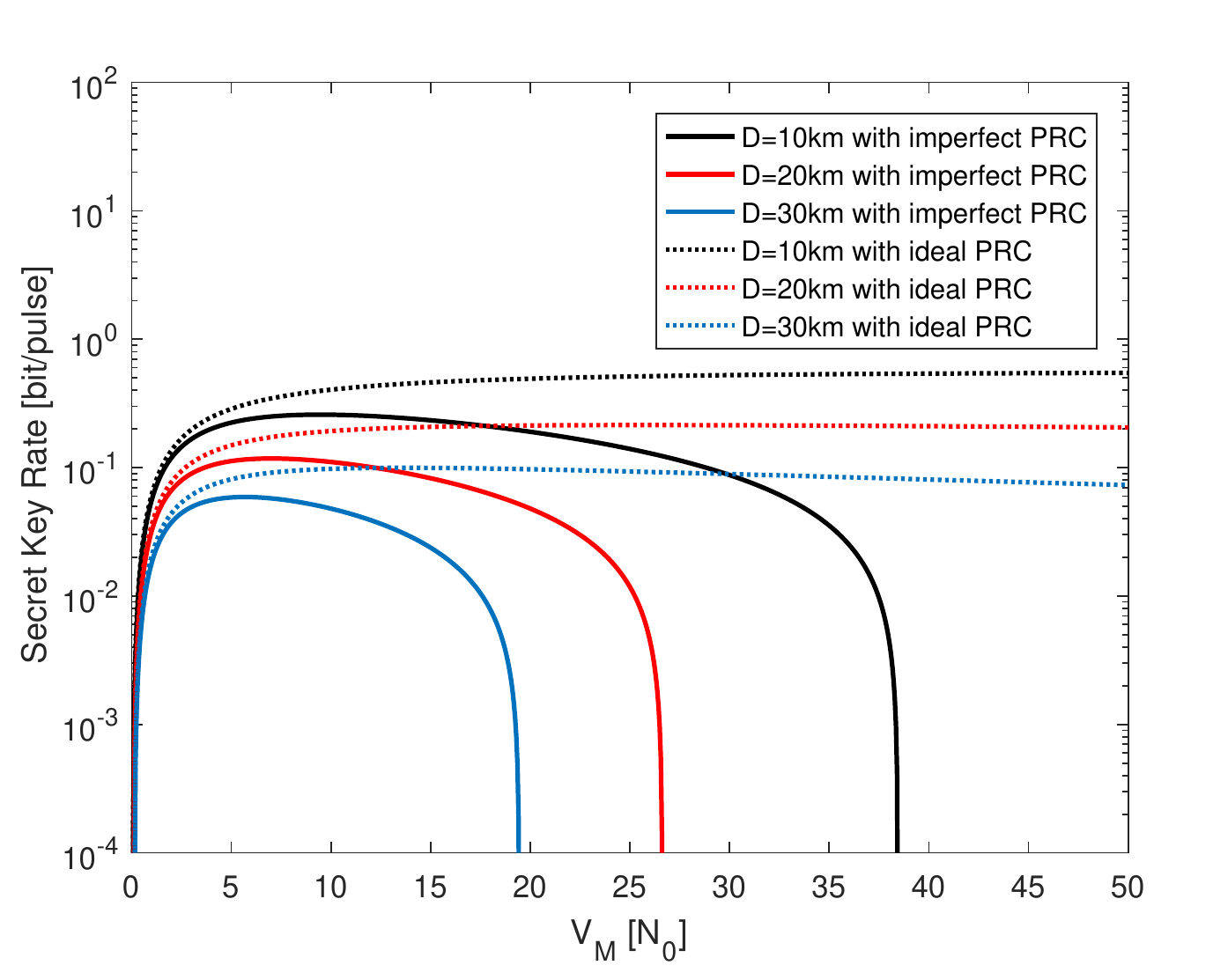}}
\caption{(Color online). Secret key rates as a function of \(V_{M}\) in the extreme asymmetric case,  where Charlie is extremely close to Bob. Transmission distances \(D=L_{AC}\) are set to 10 km, 20 km and 30 km. \(N_0\) is the shot noise variance. PRC is phase reference calibration. The solid lines denote the CV-MDI-QKD protocol with ideal phase reference calibration, the dashed lines denote the CV-MDI-QKD protocol with imperfect phase reference calibration. Parameters are fixed as follows: \(\varepsilon _A=\varepsilon _B=0.002\), \(V_{laser}=0.005\), \(|\alpha_{LO}|^2/ V_M=10^8\), reconciliation efficiency \(\beta=96\%\).}\label{fig:4}
\end{figure}
with the notations
\begin{equation}
\begin{array}{lll}
A=a^2+b^2-2c^2={V^2} + {{\eta}^2}{\left( {V + {\chi _{t}}} \right)^2} - 2\eta(V^2-1),\\ \\
B=ab-c^2= \eta(V\chi_t+1).
\end{array}
\end{equation}
The covariance matrix of the state \(\rho _{{A_1}}^{{m_{B}}}\) can be calculated as
\begin{equation}
\begin{array}{lll}
\gamma _{{A_1}}^{{m_{B}}} &=a{\rm{I}_2}-c{\sigma _z}(b{\rm{I}_2}+{\rm{I}_2})^{-1}c{\sigma _z}
\\ \\
&=[a-c^2/(b+1)]{\rm{I}_2},
\end{array}
\end{equation}
then the symplectic eigenvalues \(\lambda_3\) is given by
\begin{equation}
\lambda _3=a-c^2/(b+1)=\frac{\eta V \chi_t+V+\eta}{\eta(V+\chi_t)+1}.
\end{equation}

\section{Performance analysis}\label{PerD}
In this section, we give the numerical simulation and provide the sufficient analysis of the CV-MDI-QKD protocol with imperfect phase reference calibration compared with previous works which do not consider the impact of imperfect phase reference calibration.

\begin{figure}[!h]\center
\resizebox{8.5cm}{!}{
\includegraphics{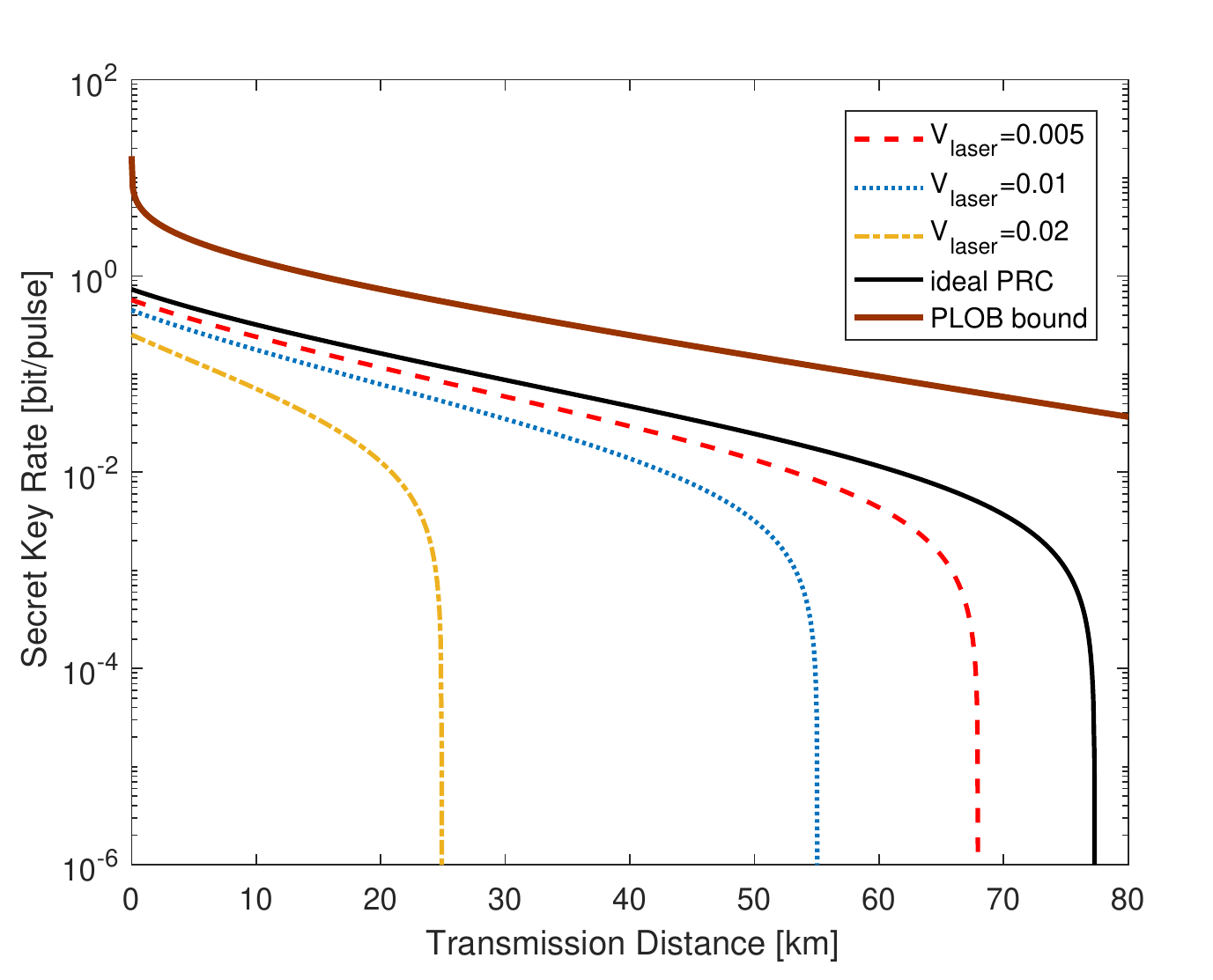}}
\label{fig5}
\caption{(Color online). Secret key rates as a function of the transmission distance in the extreme asymmetric case, where Charlie is extremely close to Bob. The uppermost heavy solid line denotes the PLOB bound. The thin solid lines denotes with the CV-MDI-QKD protocol with ideal phase reference calibration. The dashed lines denote the CV-MDI-QKD protocol with imperfect phase reference calibration, where \(V_{laser}\) are set to 0.005, 0.01 and 0.02 with the units of shot noise (\(N_0\)).Parameters are fixed as follows: \(\varepsilon _A=\varepsilon _B=0.002\), \(|\alpha_{LO}|^2/ V_M=10^8\), modulation variance \(V_M=6\), reconciliation efficiency \(\beta=96\%\).}\label{fig:5}
\end{figure}

In CV-MDI-QKD protocols, the asymmetric case, where \(L_{AC} \not= L_{BC}\) has obvious advantage in performance compared with the symmetric case, where \(L_{AC}=L_{BC}\) \cite{STN15}, and the extreme asymmetric case, where Charlie is extremely close to Bob \cite{LZY14} has the optimal performance.
In other words, the shorter the distance between Bob and Charlie, the better the performance we can obtain. On the contrary, it will degrade the performance of the system.
Employing the same parameters, the extreme asymmetric case can obtain the maximal transmission distance, which is more suitable for point-to-point communications.
In short-range network applications where the relay needs to be in the middle of the legitimate communication parties, the symmetric case is more suitable and has unique potentials.
Our following analysis is based on two cases.

\subsection{Performance analysis in the extreme asymmetric case}

The modulation variance \(V_{M}\) is critical to the performance and security of CV-MDI-QKD protocol.
Before obtaining the secret key rate of the CV-MDI-QKD protocol with imperfect phase reference calibration as a function of transmission distance in the extreme asymmetric case, we need to know how the secret key rate changes with the modulation variance in order to obtain the optimal modulation variance. We plot the secret key rates as a function of the modulation variance \(V_{M}\) with different transmission distance in the extreme asymmetric case, for both the CV-MDI-QKD protocol with ideal
phase reference calibration and the protocol with imperfect phase reference calibration, which is shown Fig.~\ref{fig:4}.
There are two key parameters, \(V_{laser}\) and \(|\alpha_{LO}|^2/ V_M\), directly decide the impact of imperfect phase reference calibration on the protocol.
\begin{figure}[!h]\center
\resizebox{8.5cm}{!}{
\includegraphics{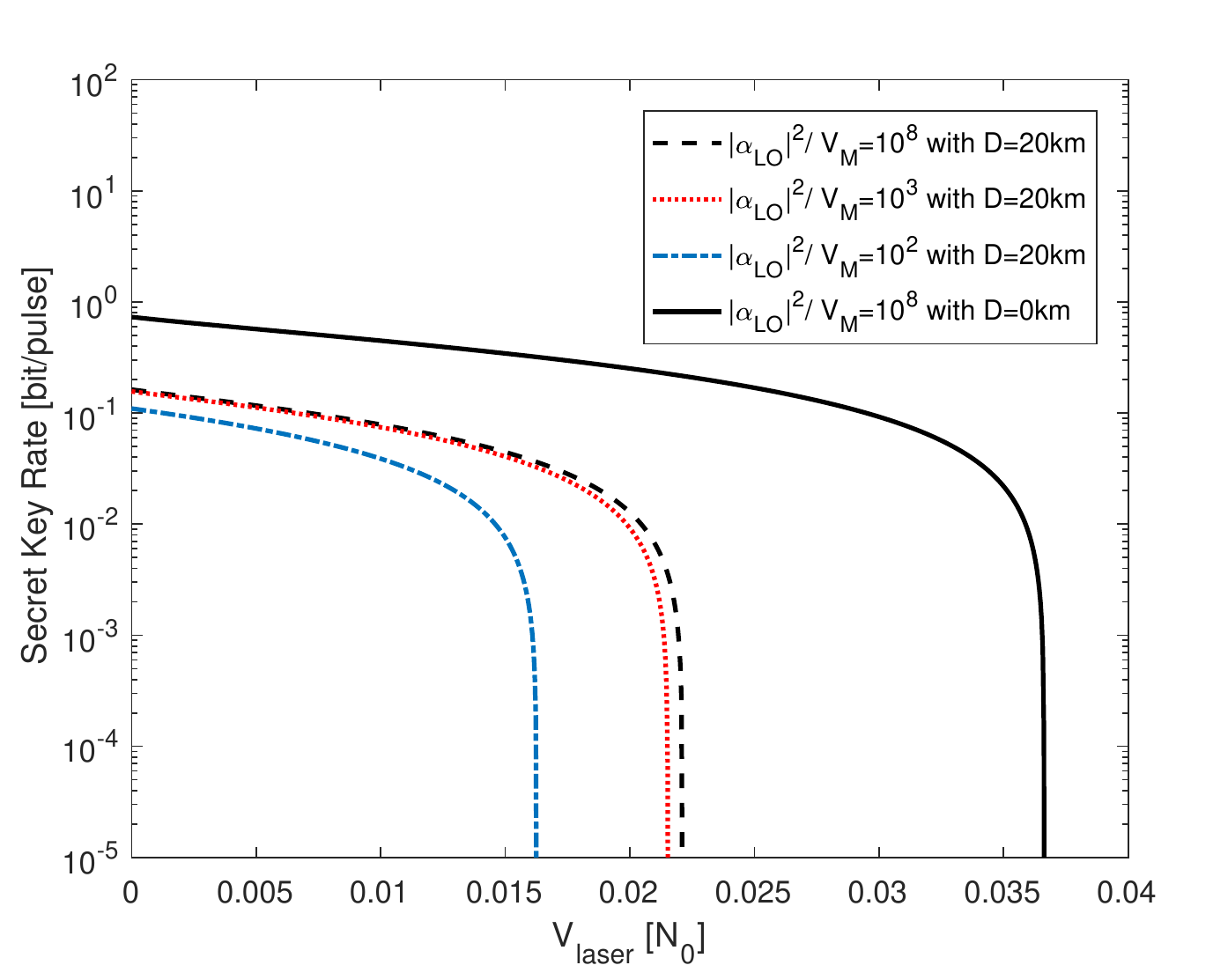}}
\caption{(Color online). Secret key rates as a function of \(V_{laser}\) in the extreme asymmetric case, where Charlie is extremely close to Bob.
The dashed lines denote the CV-MDI-QKD protocol with imperfect phase reference calibration, where transmission distances \(D=L_{AC}\) is set to 20 km and \(|\alpha_{LO}|^2/ V_M\) are set to \(10^8\), \(10^3\) and \(10^2\). The solid line denotes the initial secret key rate of
CV-MDI-QKD protocol with imperfect phase reference calibration, where transmission distances \(D=L_{AC}=\) 0 km and \(|\alpha_{LO}|^2/ V_M=10^8\).
Parameters are fixed as follows: \(\varepsilon _A=\varepsilon _B=0.002\), modulation variance \(V_M=6\), reconciliation efficiency \(\beta=96\%\).}\label{fig:6}
\end{figure}
\(V_{laser}\) is related with the spectral linewidth of two free-running lasers and the repetition rate of the system. We denote \(V_{laser}=0.005\) based on the parameters of the practical equipment.
\(|\alpha_{LO}|^2/ V_M\) is related with the light intensity of local oscillators pulse. We choose \(|\alpha_{LO}|^2/ V_M=10^8\), which is the value commonly used in practical CV systems.

Obviously, when considering the imperfection of practical phase reference calibration, the practicable \(V_{M}\) values are much lower than the one without taking this imperfection into account, which means that we need to set the modulation variance more strictly under the condition of imperfect phase reference calibration.
In addition, when transmission distance increases, the optional areas of \(V_{M}\) are gradually compressed and the secret key rate decreases evidently.
There is a noteworthy phenomenon that, under the fixed parameters, the optimal value of \(V_M\) for the CV-MDI-QKD protocol with imperfect phase reference calibration, which leads to the best performance, is always about 6 in short noise units. Hence, in the next analysis of the extreme asymmetric case, we always denote \(V_M=6\).

The plot of Fig.~\ref{fig:5} shows the secret key rates as a function of the transmission distance in the extreme asymmetric case, for both the CV-MDI-QKD protocol with imperfect phase reference calibration and the one with ideal phase reference calibration. Besides, different values of \(V_{laser}\) are taking into account for the CV-MDI-QKD protocol with imperfect phase reference calibration, and the PLOB bound is plotted as a reference for performance comparison. Here we denote \(|\alpha_{LO}|^2/ V_M=10^8\) as a fixed value.
As shown in the figure, the performance curve of the CV-MDI-QKD protocol with imperfect phase reference calibration is always lower than that of the one without considering this imperfection, and the gap between the former curve and the PLOB bound is always larger than that between the later curve and the PLOB bound. Furthermore, the gap between these two performance curves will become larger and lager with the value of \(V_{laser}\) increases, and the performance of the CV-MDI-QKD protocol with imperfect phase reference calibration reduces rapidly with \(V_{laser}\) increases.

On the one hand, these phenomena indicate that the imperfect phase reference calibration will significantly cut down the performance of the CV-MDI-QKD protocol and the reduction is more obvious with the larger \(V_{laser}\).
On the other hand, it shows that a more compact security key rate can be obtained by incorporating consideration of the imperfect phase reference calibration into security analysis.

Fig.~\ref{fig:6} depicts the secret key rates as a function of \(V_{laser}\) in the extreme asymmetric case, for the CV-MDI-QKD protocol with imperfect phase reference calibration. The lower dashed lines denote the case of secret key rate changing with \(V_{laser}\) under fixed transmission distance and different value of \(|\alpha_{LO}|^2/ V_M\).
On one hand, when the local oscillator pulse is too weak, the coherent detectors can not work effectively. On the other hand, when the local oscillator pulse is too strong, it will exceed the performance of the coherent detectors. Therefore, in the practical system, we take the intensity of local oscillator pulse as a fixed range, which leads the value of \(|\alpha_{LO}|^2/ V_M\) is always around \(10^8\).

Although the value of \(|\alpha_{LO}|^2/ V_M\) has been limited in the practical system, we still need to consider its impact on system security, as it is an important parameter in the calculation formula of \(\varepsilon_{prc}\). According to the figure, we can obtain that the value of \(|\alpha_{LO}|^2/ V_M\) and the performance of the protocol are negatively correlated. However, when \(|\alpha_{LO}|^2/ V_M\) is lager than \(10^4\), its effect on the performance of the protocol is negligible.
Therefore, in practical systems, even if the value of \(|\alpha_{LO}|^2/ V_M\) fluctuates around \(10^8\), it will not have a significant impact on the security key rate. In other words, the effect of \(|\alpha_{LO}|^2/ V_M\) on the performance of the protocol is not obvious in practice.
Hence, in the extreme asymmetric case, the most important parameter affecting the impact of imperfect phase reference calibration in practical CV-MDI-QKD systems is \(V_{laser}\).

\begin{figure}[!h]\center
\resizebox{8.5cm}{!}{
\includegraphics{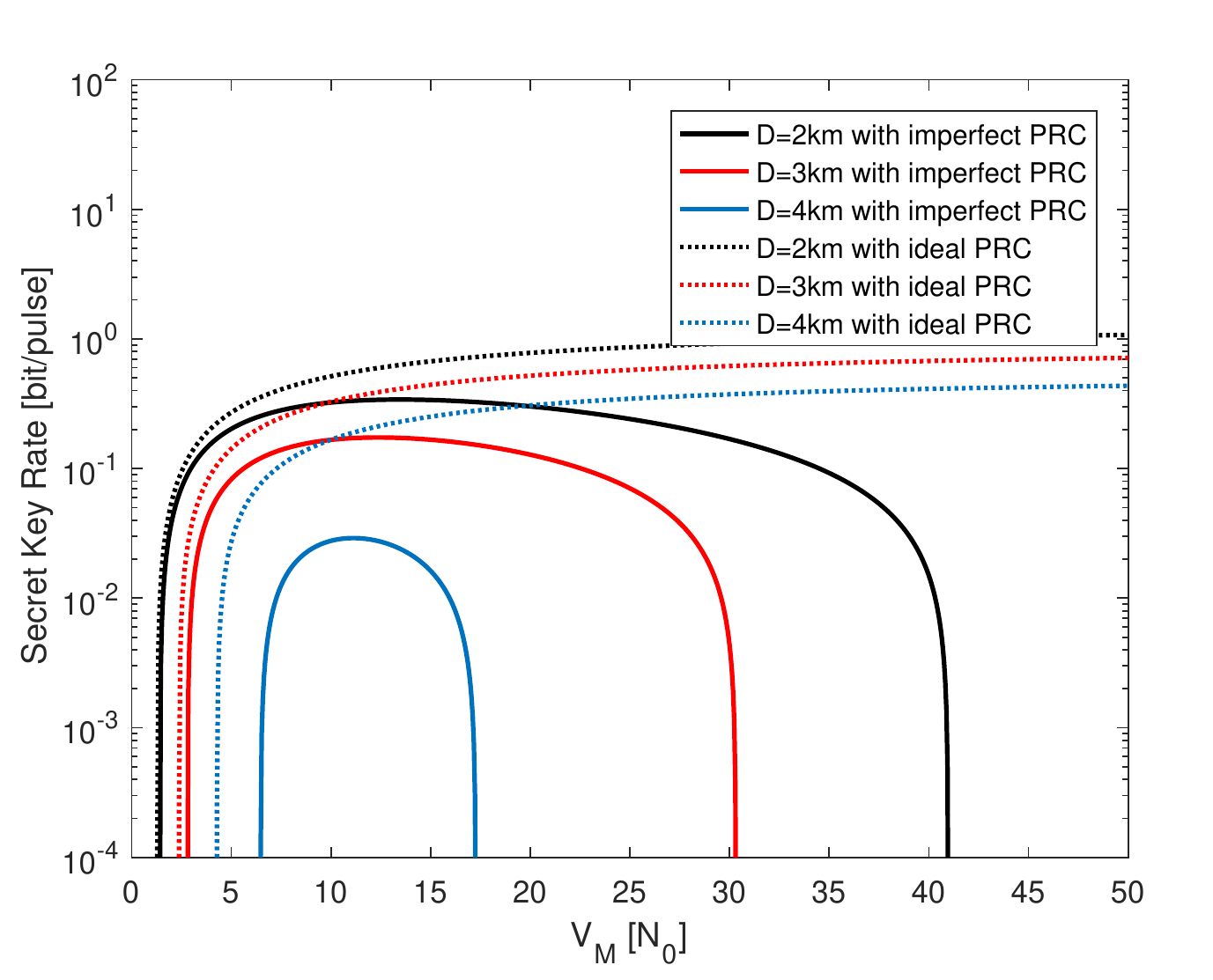}}
\caption{(Color online). Secret key rates as a function of \(V_{M}\) in the symmetric case, where Charlie is in the middle of Alice and Bob. Transmission distances \(D=L_{AC}+L_{BC}\) are set to 2 km, 3 km and 4 km. \(N_0\) is the shot noise variance. PRC is phase reference calibration. The solid lines denote the CV-MDI-QKD protocol with ideal phase reference calibration, the dashed lines denote the CV-MDI-QKD protocol with imperfect phase reference calibration. Parameters are fixed as follows: \(\varepsilon _A=\varepsilon _B=0.002\), \(V_{laser}=0.005\), \(|\alpha_{LO}|^2/ V_M=10^8\), reconciliation efficiency \(\beta=96\%\).
}\label{fig:7}
\end{figure}

The upper solid line denotes the initial secret key rate of CV-MDI-QKD protocol with imperfect phase reference calibration, where transmission distances \(D=L_{AC}=\) 0 km and \(|\alpha_{LO}|^2/ V_M=10^8\).
This line shows the tolerance of the CV-MDI-QKD protocol to \(V_{laser}\) when the CV-MDI-QKD system works properly.
Under the fixed parameters, the CV-MDI-QKD protocol will have no secret key rate when \(V_{laser}\) lager than 0.03665.
In other words, in the extreme asymmetric case, the upper limit of system tolerance to \(V_{laser}\) is 0.03663, and the tolerance of the CV-MDI-QKD protocol to the excess noise \(\varepsilon_{prc}\) introduced by imperfect phase reference
calibration is obtained as about 0.03663\(V_M\) in short noise units.

\subsection{Performance analysis in the symmetric case}

In the symmetric case, the untrusted third part Charlie is right in the middle of Alice and Bob, which is quite suitable for the applications where two legitimate parties are roughly equidistant from a public server.
Same as the previous subsection, we should obtain the optimal value of \(V_M\) before  simulating the secret key rate of the CV-MDI-QKD protocol with imperfect phase reference calibration in the symmetric case.
The plot of Fig.~\ref{fig:7} shows the secret key rates as a function of the modulation variance \(V_{M}\) with different transmission distance in the symmetric case, for both the CV-MDI-QKD protocol with ideal
phase reference calibration and the protocol with imperfect phase reference calibration. The feasible range of \(V_{M}\) in the latter is much smaller than that in the former.
Considering the imperfection of phase reference calibration, with transmission distance increases, the optional areas of \(V_{M}\) are gradually compressed, which is similar to what is shown in Fig.~\ref{fig:4}. Under the fixed parameters, the optimal value of \(V_M\) in the symmetric case is about 12.

\begin{figure}[!h]\center
\resizebox{8.5cm}{!}{
\includegraphics{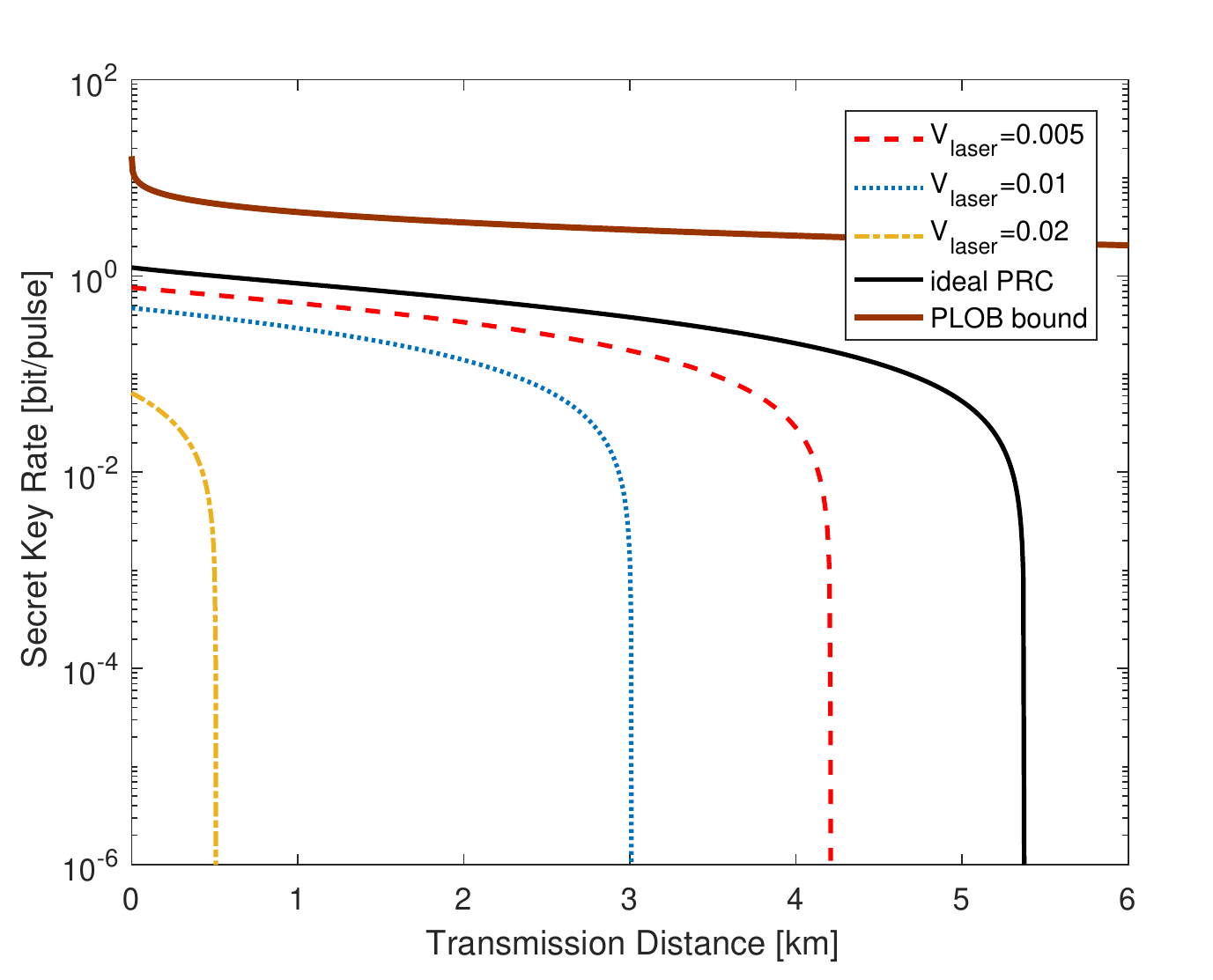}}
\label{fig5}
\caption{(Color online). Secret key rates as a function of the transmission distance in the symmetric case, where Charlie is in the middle of Alice and Bob. The uppermost heavy solid line denotes the PLOB bound. The thin solid lines denotes with the CV-MDI-QKD protocol with ideal phase reference calibration. The dashed lines denote the CV-MDI-QKD protocol with imperfect phase reference calibration, where \(V_{laser}\) are set to 0.005, 0.01 and 0.02 with the units of shot noise (\(N_0\)).Parameters are fixed as follows: \(\varepsilon _A=\varepsilon _B=0.002\), \(|\alpha_{LO}|^2/ V_M=10^8\), modulation variance \(V_M=12\), reconciliation efficiency \(\beta=96\%\). }\label{fig:8}
\end{figure}

\begin{figure}[!h]\center
\resizebox{8.5cm}{!}{
\includegraphics{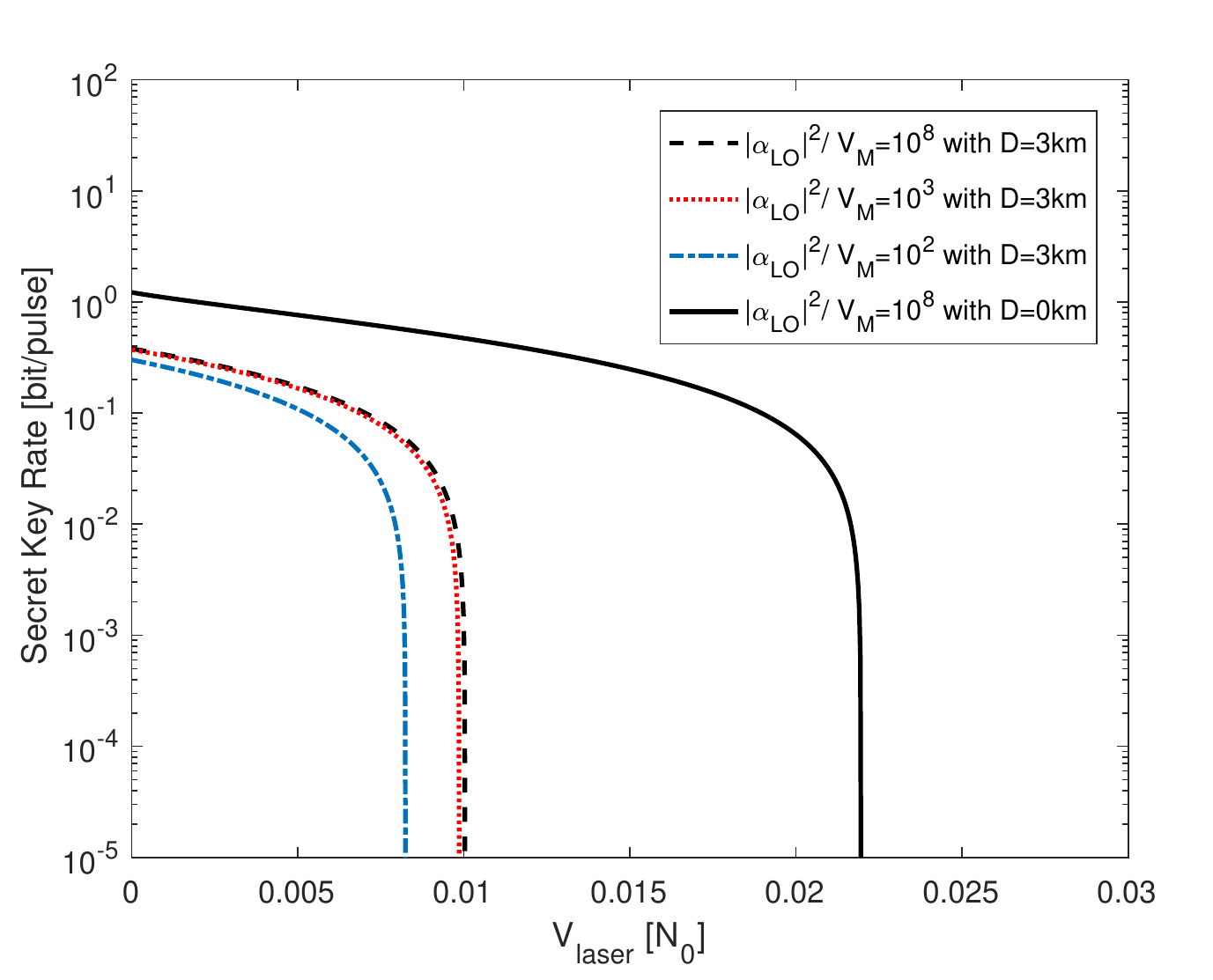}}
\label{fig5}
\caption{(Color online). Secret key rates as a function of \(V_{laser}\) in the symmetric case, where Charlie is in the middle of Alice and Bob.
The dashed lines denote the CV-MDI-QKD protocol with imperfect phase reference calibration, where transmission distances \(D=L_{AC}+L_{BC}\) is set to 3 km and \(|\alpha_{LO}|^2/ V_M\) are set to \(10^8\), \(10^3\) and \(10^2\). The solid lines denote the initial secret key rate of
CV-MDI-QKD protocol with imperfect phase reference calibration, where transmission distances \(D=L_{AC}=\) 0 km and \(|\alpha_{LO}|^2/ V_M=10^8\).
Parameters are fixed as follows: \(\varepsilon _A=\varepsilon _B=0.002\), modulation variance \(V_M=12\), reconciliation efficiency \(\beta=96\%\).}\label{fig:9}
\end{figure}

The plot of Fig.~\ref{fig:8} shows the secret key rates as a function of the transmission distance in the symmetric case, for both the CV-MDI-QKD protocol with imperfect phase reference calibration and the one with ideal phase reference calibration.
The modulation variance \(V_M\) of both protocols are all set to 12,  \(|\alpha_{LO}|^2/ V_M\) is also fixed as \(10^8\).
Same as the analysis of Fig.~\ref{fig:5}, the performance of the CV-MDI-QKD protocol with imperfect phase reference calibration is always worse than that of the one without considering this imperfection, and the gap will become lager rapidly with \(V_{laser}\) increases. Furthermore, for both the CV-MDI-QKD protocol with imperfect phase reference calibration and the one with ideal phase reference calibration, the maximal transmission distances of the symmetric case are less than a tenth of these of the extreme asymmetric case. The secret key rate of the CV-MDI-QKD protocol with imperfect phase reference calibration in the symmetric case looks more sensitive to the change of \(V_{laser}\) than that in the extreme asymmetric case, which will be confirmed in Fig.~\ref{fig:9}.

Fig.~\ref{fig:9} depicts the secret key rates of the CV-MDI-QKD protocol with imperfect phase reference calibration as a function of \(V_{laser}\) in the symmetric case, with different values of \(|\alpha_{LO}|^2/ V_M\) and transmission distance.
Similar to what is shown in Fig.~\ref{fig:6}, although \(|\alpha_{LO}|^2/ V_M\) and the performance of the protocol have negative correlation, when \(|\alpha_{LO}|^2/ V_M\) surpasses \(10^4\), its effect on the performance of the protocol is not worth mentioning. So the most critical parameter for determining the impact of the imperfect phase reference calibration in practical CV-MDI-QKD systems is still \(V_{laser}\) in the symmetric case.
The upper solid line shows the tolerance of the CV-MDI-QKD protocol to \(V_{laser}\) in the symmetric case, where the the upper limit of \(V_{laser}\) is 0.0220. It shows that the secret key rate of the CV-MDI-QKD protocol with imperfect phase reference calibration in the symmetric case looks more sensitive to \(V_{laser}\) than that in the extreme asymmetric case.
Then, the tolerance of the CV-MDI-QKD protocol to \(\varepsilon_{prc}\) can be calculated as about 0.0220\(V_M\) in short noise units.

~\\
\section{Conclusion and Discussions}\label{Con} 

In this paper, we have investigated the imperfection of practical phase reference calibration operation on the security of CV-MDI-QKD protocol, which is caused by the non-synchronization of two remote lasers in senders and has not been taken into account in previous security analysis of this protocol.
We developed a comprehensive security framework to model and characterize this imperfection.
Through reasonable modeling, the effect of this imperfection on the security of the CV-MDI-QKD protocol is equivalent to the excess noise \(\varepsilon_{prc}\) introduced by imperfect phase reference calibration. A tight bound of the security key rate is derived under arbitrary collective attacks.
The qualitative and quantitative security analysis shows that the imperfect phase reference calibration will damage the performance and security of the CV-MDI-QKD protocol.
This work will get ride of the security hazards led by the imperfect phase reference calibration without the adjustment of the protocol structure.

In the analysis of \(\varepsilon_{prc}\), we find that the most critical parameter for determining the impact of the imperfect phase reference calibration in practical CV-MDI-QKD systems is \(V_{laser}\), which is a fixed parameter in the specific system and decided by the spectral linewidth of two free-running lasers and the repetition rate \(f\) of the system. We usually choose \(f\) below 100 MHz with considing the current bandwidth limitation of shot-noise limited coherent detectors. In order to minimize \(V_{laser}\), we can choose low-phase-noise lasers, such as external-cavity lasers (ECL), whose typical spectral linewidth is of a few kHz \cite{SFCV}. In this case, \(V_{laser}\) may even be less than \(10^{-4}\).
The participation of such equipment can effectively narrow the impact of the imperfect phase reference calibration on the security and performance of CV-MDI-QKD protocol.
In future work, we will strive to design a comprehensive security architecture to characterize the overall practical security of CV-MDI-QKD protocol.
\\ \\
\begin{acknowledgments}
This work was supported the National key research and development program (Grants No. 2016YFA0302600), the National Basic Research Program of China (Grant No. 2013CB338002) and the National Natural Science Foundation of China (Grants No. 11304397, 61332019, 61505261, 61675235, 61605248, 61671287).
\end{acknowledgments}

\end{document}